\documentclass[noshowpacs,superscriptaddress,amsmath,amssymb]{revtex4}
\usepackage{xcolor}
\usepackage{graphicx}% Include figure files
\usepackage{dcolumn}% Align table columns on decimal point
\usepackage{bm}% bold math

\usepackage[cp1251]{inputenc}

\def\bfg #1{{\mbox{\boldmath $#1$}}}
\begin{document}
%\selectlanguage{russian} % Для  статьи на русском языке
%\selectlanguage{english} % Для статьи на английском языке
%
%\title{Эффект нарушения $T$-инвариантности в $\bm{dd}$-рассеянии}
%
%\author{М.Н. Платонова}
%
%\title{$T$- invariance violation effect in  $\bm{dd}$-scattering}
\title{{{Time-reversal}} invariance violation effect in $\bm{dd}$ scattering \footnote{This research was carried out at the expense of
the grant from the Russian Science Foundation
(project No. 23-22-00123,
https://rscf.ru/en/project/23-22-00123/)}
}

\author{M.N. Platonova}
\email[]{platonova@nucl-th.sinp.msu.ru}
%\homepage[]{Your web page}

\affiliation{
%Skobeltsyn Institute of Nuclear Physics, Moscow State University, Moscow, %119991 Russia
Skobeltsyn Institute of Nuclear Physics, Lomonosov Moscow State University, \\ Leninskie Gory 1/2, 119991 Moscow, Russia
}
\affiliation{
%Dzhelepov Laboratory of Nuclear Problems, Joint Institute for Nuclear %Research, Dubna, Moscow region, 141980 Russia
Dzhelepov Laboratory of Nuclear Problems, Joint Institute for Nuclear Research, \\ Joliot-Curie 6, 141980 Dubna, Moscow region, Russia
}
\author{Yu.N. Uzikov}
\email[]{uzikov@jinr.ru}
%\homepage[]{Your web page}
%\thanks{}
\affiliation{
Dzhelepov Laboratory of Nuclear Problems, Joint Institute for Nuclear Research, \\ Joliot-Curie 6, 141980 Dubna, Moscow region, Russia
}
\altaffiliation[Also ]{
%Государственный университет ``Дубна'', Дубна,
%ул. Университетская 19,
%141980, Россия
Dubna State University, Universitetskaya 19, 141980 Dubna, Moscow region, Russia
}
%\affiliation{Лаборатория ядерных проблем им. В.П. Джелепова, %Объединенный институт ядерных исследований, Дубна,
%ул. Жолио-Кюри, 6,
%%141980 Россия}
\affiliation{
Faculty of Physics, Lomonosov  Moscow State University, Leninskie Gory 1, 119991 Moscow, Russia
}

%\date{\today}% It is always \today, today,
             %  but any date may be explicitly specified

\begin{abstract}
A formalism has been developed for calculating the signal of violation of
time-reversal invariance, provided that
space-reflection ({parity}) invariance
is conserved during the scattering of tensor-polarized deuterons on vector-polarized
 ones. { The formalism is} based on the Glauber theory with the full consideration of spin dependence of $NN$ elastic scattering amplitudes and spin structure
of colliding deuterons.
The numerical calculations have been carried out in the range of laboratory proton energies of $T_p=0.1$--$1.2$ GeV using the SAID database
for spin amplitudes and in the energy region of the SPD NICA experiment corresponding
to the invariant mass of the interacting nucleon pairs $\sqrt{s_{NN}}= 2.5$--$25$ GeV, using two phenomenological models of $pN$ elastic scattering.
It is found that only one type of the { time-reversal} non-invariant
{ parity} conserving
 $NN$ interaction gives a non-zero contribution to the signal
 in question, that is important for isolating an unknown constant of this interaction from the corresponding data.

%PACS numbers may be entered using the \verb+\pacs{#1}+ command.
\end{abstract}

%\pacs{Valid PACS appear here \verb+\pacs{#1}+ 21.80.+a}% PACS, the Physics and Astronomy Classification Scheme.

%Use showkeys class option if keyword display desired

\maketitle

%\section{\label{sec:level1}First-level heading:\protect\\ The line
%break was forced \lowercase{via} \textbackslash\textbackslash}

%%%%%%%%%%%%%%%%%%%%%%%%%%%%%%%%%%%%%%%%%%%  DSPIN-2023 %%%%%%%%%%%%%%%%%%%%%%%%%%%%%%%%%%%%%%%%%%%%%%%%
\section{\label{sec1} Introduction}

Discrete symmetries with respect to the time reversal ($T$), space reflection ($P$) and charge
conjugation ($C$) play a key role in theory of fundamental interactions and astrophysics.
Under $CPT$-symmetry which takes place in local quantum field theory
 \cite{Luders:1954zz,Pauli:1955},
the violation of $T$-invariance means also the violation of $CP$-symmetry,
which is necessary to explain the baryon asymmetry of the Universe \cite{Sakharov:1967dj}.
  $CP$ violation observed in decays of $K$, $B$ and $D$ mesons is consistent with the standard model (SM) of fundamental interactions, but it turns out to be far from sufficient to explain the observed baryon asymmetry
 \cite{Riotto:1999yt}.
 This means that there must be other sources of $CP$ violation in nature beyond the SM.

One of these sources is associated with the electric dipole moments (EDM) of free elementary particles, neutral atoms and the lightest nuclei, the search for which is given great attention during the last decades
{ \cite{Chupp:2017rkp}}.
%
%\cite{CPEDM:2019nwp}.
An observation of a non-zero EDM value will mean
that $T$-invariance and { parity} are violated simultaneously.
%  In neutrino physics, $CP$ violation, which is also actively searched for during the last decade in %experiments, would also mean simultaneous $T$ and $P$ violation.
%
{
 Much less attention was paid to the experiments on the search for the effects of $T$-invariance
 violation with { parity} conserved (TVPC) and flavor conserved.
This type of interaction was introduced  in
\cite{Okun:1965tu} to explain the CP violation observed in
 kaons decays
and is related to physics outside  of the SM
\cite{Gudkov:1991qg,Vergeles:2022mqu}}.
%and TVPC effects can only manifest themselves
% due to weak $P$-odd radiation corrections
%to the known $T$-odd $P$-odd effects
%\cite{Gudkov:1991qg},
% being in such a case too small for experimental detection.
% \cite{Conti:1992xn,Engel:1995vv}.
% However, beyond the SM, the coupling constant
% of the TVPC interaction may be much stronger \cite{Gudkov:%1991qg,Vergeles:2022mqu}}.
%
  As was shown
  in a model-independent way
  within the effective field theory \cite{Kurylov:2000ub},
 due to an unknown mechanism of EDM generation, the available experimental limitations on EDM
 cannot be used for
 estimation of the appropriate restrictions on TVPC effects, the detection of which at the current level of experimental
sensitivity would be the direct evidence of physics beyond the SM.

In the scattering of two polarized nuclei, the signal of violation of $T$-invariance while conserving { parity} is
that
   { the}
   component of the total cross section which corresponds to the interaction of a transversely polarized ($P_y$) incident nucleus with a tensor polarized
($P_{xz}$) target nucleus
  \cite{Barabanov:1986sz}.
This observable cannot be simulated by the interaction in the initial or final states and is not zero only in the presence of the TVPC interaction discussed here, in a similar way as
 EDM is a signal of a $T$- { and} $P$-violating interaction.

%%YU %%%%%%%%%%%%%%%%%%%%%%%%%

{ Following  the
%was shown in
% justification
  description of the experimental COSY project
for studying the TVPC effect in
$pd$ interactions \cite{Lenisa:2019cgb}, this type of component of the total cross section (known
in the literature as the TVPC null-test signal), can also be measured  in $dd$ scattering}
by measuring the asymmetry of the event counting rate
in this process. This asymmetry appears,
  when the sign of the vector polarization
of one of the colliding deuterons ($P_y^{(1)}$) is changed, whereas the tensor polarization ($P_{xz}^{(2)}$) of the second deuteron is unchanged.

%%% OLD%%%%%%%%%%%%%%%%%%%%%%%%%%%%%
%A measurement of
%such component of the total cross section (called
%in the literature the TVPC null-test signal),
%as shown in
%% justification
%{\color{red}  description} of the experimental COSY project
%for
% study TVPC in
%$pd$ interaction \cite{Lenisa:2019cgb},
%is   possible {\color{red} also for $dd$} scattering
%by measuring the asymmetry of the event counting rate
%in {\color{red} the process},
%  when the sign of the vector polarization
%of one of the colliding deuteron ($P_y^{(1)}$) changes while the tensor %polarization ($P_{xz}^{(2)}$) of the second deuteron is not changed.
%%%%%%%%% END of OLD
When using this method, the transverse vector polarization $P_y^{(2)}$ of the second (tensor polarized) deuteron is required to be zero \cite{Temerbayev:2015foa}.
Another method of measurement which does not require such a restriction on $P_y^{(2)}$, but
uses the rotating polarization of the incoming beam
 in a combination with the Fourier analysis of the time-dependent counting rate of the number of events, was proposed
 in Ref. \cite{Nikolaev:2020wsj}.
 A possible measurement procedure of the TVPC null-test signal in $dd$ scattering was recently briefly discussed in~\cite{Uzikov:2024}.

 %%%YU
 Here we focus on the
 theoretical calculation of the TVPC null-test signal.
 Its dependence on the collision energy
 for $pd$
\cite{Uzikov:2015aua, Uzikov:2016lsc} and
$^3{\rm He}d$ \cite{Uzikov:2023eex} scattering
  was performed
%  recently
in the Glauber theory in the range of laboratory energies 0.1--1 GeV, taking into account the full spin dependence of $NN$ scattering amplitudes as well as $S$ and $D$ components of the deuteron wave function.
%%

%The theoretical calculation of the TVPC null-test signal dependence on the collision energy in the %range of laboratory energies 0.1 -- 1 GeV was performed recently
%in the Glauber theory, taking into account the full spin dependence of $NN$ scattering amplitudes as %well as $S$ and $D$ components of the deuteron wave function, for $pd$
%\cite{Uzikov:2015aua, Uzikov:2016lsc} and
%$^3{\rm He}d$ \cite{Uzikov:2023eex} scattering.

In this paper, we calculate the TVPC null-test signal in $dd$ scattering for the first time using the fully spin-dependent Glauber theory
%%%YU
 for this process
  and generalizing the method
  developed  previously in \cite{Uzikov:2024, Uzikov:2015aua, Uzikov:2016lsc}.
%  In this paper, we calculate the TVPC null-test %signal in $dd$ scattering for the first time using %the fully spin-dependent Glauber theory and %generalizing the method
%  developed  previously in \cite{Uzikov:2015aua, %Uzikov:2016lsc}.
    The following section~\ref{sec2} provides the basic mathematical formalism for this calculation. In Sec.~\ref{sec3}, the results of numerical calculations are presented and analysed.
    Conclusions are given in Sec.~\ref{sec4}.
    In Appendix, a detailed derivation of the final formulas for the TVPC signal is given.

%\section{\label{sec2} ВЫЧИСЛЕНИЕ TVPC СИГНАЛА В $dd$-РАССЕЯНИИ}
\section{\label{sec2} Calculation of the TVPC signal in $\bm{dd}$ scattering}

 Considering $dd$ scattering, one should note that
   in $pd$ collision, the TVPC signal is determined by the component of the total cross section corresponding to a vector-polarized proton interacting with a tensor-polarized deuteron~\cite{Nikolaev:2020wsj}. Unlike
$pd$, the $dd$ scattering has two symmetric components of the total cross section corresponding to the vector polarization of one deuteron and the tensor polarization of the other.
Accordingly, the TVPC transition operator $dd \to dd$ at zero angle includes two terms:
\begin{equation}
\label{mtv}
\hat M_{\rm TVPC}(0) = g_1 \hat O_1 + g_2 \hat O_2.
\end{equation}
Here the operators $\hat O_1$ and $\hat O_2$ are defined as
\begin{eqnarray}
\label{o12}
\hat O_1 &=& \hat k_m \hat Q^{(1)}_{mn} \varepsilon_{nlr} S^{(2)}_l \hat k_r, \nonumber \\
\hat O_2 &=& \hat k_m \hat Q^{(2)}_{mn} \varepsilon_{nlr} S^{(1)}_l \hat k_r,
\end{eqnarray}
where ${\hat {\bf k}}$ is
a unit  vector directed along the incident beam,
$S^{(j)}_l$
%%YU
%%($l=x,y,z$)
 are the components of the spin operator of the $j$-th deuteron,
$\hat Q^{(j)}_{mn} = \frac{1}{2}\left(S^{(j)}_m S^{(j)}_n + S^{(j)}_n S^{(j)}_m - \frac{4}{3}\delta_{mn}I\right)$  is
the symmetric tensor operator, and
%%YU
{ $\varepsilon_{nlr}$ is the fully antisymmetric tensor ($m,n,l,r=x,y,z$)}.
Here and further, we assume $j = 1$ for the incident deuteron and $j = 2$ for the target one.

We find the TVPC signal using the optical theorem:
\begin{eqnarray}
\label{sigtv}
\sigma_{\rm TVPC} &=& 4 \sqrt{\pi} {\rm Im \, Tr}(\hat \rho_i \hat M_{\rm TVPC}(0)) = \sigma^{(1)}_{\rm TVPC} + \sigma^{(2)}_{\rm TVPC},
\end{eqnarray}
where $\hat \rho_i$ is
the spin density matrix of the initial state,
which
includes vector and tensor polarizations of both deuterons,
and the cross sections $\sigma^{(i)}_{\rm TVPC}$ ($i=1,2$)
are expressed through the amplitudes $g_i$ as follows:
\begin{eqnarray}
\sigma^{(1)}_{\rm TVPC} &=& 4\sqrt{\pi} {\rm Im}\left(\frac{g_1}{9}\right)(P^{(1)}_{xz}P^{(2)}_y - P^{(1)}_{zy}P^{(2)}_x), \nonumber \\
\sigma^{(2)}_{\rm TVPC} &=& 4\sqrt{\pi} {\rm Im}\left(\frac{g_2}{9}\right)(P^{(2)}_{xz}P^{(1)}_y - P^{(2)}_{zy}P^{(1)}_x).
\label{opticalt}
\end{eqnarray}
In turn, the amplitudes $g_1$ and $g_2$ can be expressed in terms of matrix elements from the transition operator over the spin states of the incident and target deuterons in the initial and final states,
$<m_1',m_2'|\hat M_{\rm TVPC}(0)|m_1,m_2>$:
\begin{eqnarray}
\label{g12m}
<-1,1|\hat M_{\rm TVPC}(0)|0,0> = i\frac{g_1 + g_2}{2}, \nonumber \\
<1,0|\hat M_{\rm TVPC}(0)|0,1> = i\frac{g_1 - g_2}{2}.
\end{eqnarray}

Let's find the transition operator $\hat M_{\rm TVPC}(0)$ in the Glauber model, taking spin effects into account. A single-scattering mechanism, as well as in the case of $pd$ collisions, does not contribute to the TVPC signal,
since the corresponding TVPC $NN$ amplitude
is vanishing at the zero scattering angle~\cite{Temerbayev:2015foa}. In this paper, we calculate the TVPC signal in the double-scattering approximation, neglecting the contributions of triple and quadruple $NN$ collisions, which give only a small correction to the $dd$ elastic differential cross section at forward scattering angles~\cite{Alberi:1981af,Uzikov:2024}.

The amplitude of the double-scattering mechanism in an elastic $dd$ collision consists of two terms,
the so-called ``normal'' and ``abnormal'' ones.
The first
%%MP of them -OK
(``normal'') corresponds to the sequential scattering of both nucleons of the
incident deuteron on one of the nucleons of the target deuteron,
and similarly,
of one of the nucleons in the incident beam on both nucleons of the target.
The second (``abnormal'') is the simultaneous collision of one nucleon from the incident beam
with one of the target nucleons and another nucleon of the beam with another nucleon of
the target. The corresponding scattering amplitude at zero angle takes the form:
\begin{widetext}
\begin{eqnarray}
\label{mg2}
\hat M^{(2)}(0) &=& \hat M^{(2n)}(0) + \hat M^{(2a)}(0), \nonumber \\
\hat M^{(2n)}(0) &=& \frac{i}{2\pi^{3/2}}\int \int \int d^3 \rho d^3 r d^2 q \Psi^+_{d(12)}({\bf r}) \Psi^+_{d(34)}(\bm\rho) \left[e^{i{\bf q} \bm\delta}\hat O^{(2n)}({\bf q}) + e^{i\bf{qs}}\hat O'^{(2n)}({\bf q})\right] \Psi_{d(34)}(\bm\rho) \Psi_{d(12)}({\bf r}), \nonumber \\
\hat M^{(2a)}(0) &=& \frac{i}{2\pi^{3/2}}\int \int \int d^3 \rho d^3 r d^2 q \Psi^+_{d(12)}({\bf r}) \Psi^+_{d(34)}(\bm\rho) e^{i{\bf q}({\bf s}-\bm\delta)}
\left[\hat O^{(2a)}({\bf q}) + \hat O'^{(2a)}({\bf q})\right]\Psi_{d(34)}(\bm\rho) \Psi_{d(12)}({\bf r}).
\end{eqnarray}
\end{widetext}
The operators
$\hat O^{(2n)}({\bf q})$, $\hat O'^{(2n)}({\bf q})$,
$\hat O^{(2a)}({\bf q})$ and  $\hat O'^{(2a)}({\bf q})$
are expressed in terms of spin-dependent $NN$ amplitudes:
\begin{eqnarray}
\label{og2}
\hat O^{(2n)}({\bf q}) &=& \frac{1}{2} \{M_{31}({\bf q}),M_{41}(-{\bf q})\} \nonumber \\
&&+ \frac{1}{2} \{M_{32}({\bf q}),M_{42}(-{\bf q})\}, \nonumber \\
\hat O'^{(2n)}({\bf q}) &=& \frac{1}{2} \{M_{31}({\bf q}),M_{32}(-{\bf q})\} \nonumber \\
&&+ \frac{1}{2} \{M_{41}({\bf q}),M_{42}(-{\bf q})\}, \nonumber \\
\hat O^{(2a)}({\bf q}) &=& M_{31}({\bf q}) M_{42}(-{\bf q}), \nonumber \\
\hat O'^{(2a)}({\bf q}) &=& M_{32}({\bf q}) M_{41}(-{\bf q}).
\end{eqnarray}
Here, the subscripts 1 and 2 refer to the nucleons of the target deuteron, and 3 and 4 refer to the nucleons of the incoming deuteron;
${\bf r} = {\bf r}_1 - {\bf r}_2$,  $\bm{\rho} = {\bf r}_3 - {\bf r}_4$;
${\bf s}$ and
$\bm{\delta}$
are the components of the vectors ${\bf r}$ and $\bm{\rho}$, respectively, perpendicular to the direction of the incident beam.
In the Glauber approximation, one can put
${\bf qr} = {\bf qs}$ and
${\bf q}\bm{\rho} = {\bf q}\bm\delta$.

The deuteron wave function is represented in a standard way:
\begin{equation}
\label{psid} \Psi^d_{(ij)} = \frac{1}{\sqrt{4\pi}r}\left(u(r) + \frac{1}{2\sqrt{2}}w(r)\hat S_{12}(\hat{\bf r};{\bm \sigma}_i,{\bm \sigma}_j)\right),
\end{equation}
 where $u(r)$ and  $w(r)$ are the $S$- and $D$-wave radial functions, $\hat S_{12}(\hat {\bf r};{\bm \sigma}_i,{\bm \sigma}_j) = 3({\bm \sigma}_i\cdot\hat r)({\bm \sigma}_j\cdot\hat r) - {\bm \sigma}_i\cdot{\bm \sigma}_j$
 is { the} tensor operator,
and  $\frac{1}{2}{\bm \sigma}_i$ -- the spin operator of the $i$th nucleon.

For $T$-even $P$-even $NN$ amplitudes, we
use the following representation~\cite{Platonova:2010wjt}:
\begin{eqnarray}
\label{mpn}
 M_{ij}({\bf q}) &=& A_N + C_N({\bfg \sigma}_i\cdot\hat{\bf n}) + C'_N({\bfg \sigma}_j\cdot\hat{\bf n}) \nonumber \\
 && + B_N({\bfg \sigma}_i\cdot\hat{\bf k})({\bfg \sigma}_j\cdot\hat{\bf k})  \nonumber \\
 && + (G_N + H_N)({\bfg \sigma}_i\cdot\hat{\bf q})({\bfg \sigma}_j\cdot\hat{\bf q})  \nonumber \\
 && + (G_N - H_N)({\bfg \sigma}_i\cdot\hat{\bf n})({\bfg \sigma}_j\cdot\hat{\bf n}).
\end{eqnarray}
Here the unit vectors  $\hat{\bf k}, \hat{\bf q}, \hat{\bf n}$ correspond to the vectors
\begin{equation}
\label{eq-kqn}
 {\bf k} = \frac{1}{2}({\bf p}+{\bf p}'), \quad {\bf q} = {\bf p}-{\bf p}', \quad
 {\bf n} = [{\bf p}' \times {\bf p}],
\end{equation}
${\bf p}$ and ${\bf p}'$
are the momenta of the incident and scattered nucleon, and the invariant amplitudes
$A_N, C_N, C'_N, B_N, G_N, H_N$ ($N = p$ for $\{ij\} = \{31\}, \{42\}$ and $N = n$  for $\{ij\} = \{32\}, \{41\}$) depend on the momentum $q = |{\bf q}|$.
To calculate $M_{ij}(-{\bf q})$, one should replace  ${\bf q} \to -{\bf q}$, ${\bf n} \to -{\bf n}$ in Eq.~(\ref{mpn}).
In the laboratory frame traditionally used to derive scattering amplitudes in the Glauber model, the amplitudes $C_N$ and $C'_N$ are different.

The amplitudes~(\ref{mpn}) are normalized in such a way that
\begin{equation}
\frac{d\sigma_{ij}}{dt} = \frac{1}{4}Tr(M_{ij} M^{+}_{ij}).
\end{equation}
In turn, the amplitudes of $dd$ elastic scattering are related to the differential cross section as
follows:
\begin{equation}
 \frac{d\sigma}{dt}=\frac{1}{9}Tr(\hat M \hat M^{+}).
\end{equation}
This relation is consistent with the optical theorem (\ref{opticalt}).

Further, we take the TVPC $NN \to NN$ transition operator in the form~\cite{Temerbayev:2015foa}:
\begin{eqnarray}
\label{tpn}
t_{ij} &=& h_N[({\bfg \sigma}_i \cdot {\bf k})({\bfg \sigma}_j \cdot {\bf q}) +
({\bfg \sigma}_i \cdot {\bf q})({\bfg \sigma}_j \cdot {\bf k}) \nonumber \\
&& - \frac{2}{3}({\bfg \sigma}_i \cdot {\bfg \sigma}_j)({\bf q} \cdot {\bf k})]/m^2 \nonumber \\
&& + g_N[{\bfg \sigma}_i \times {\bfg \sigma}_j]\cdot[{\bf q}\times {\bf k}]({\bfg \tau}_i - {\bfg \tau}_j)_z/m^2 \nonumber \\
&& + g'_N({\bfg \sigma}_i - {\bfg \sigma}_j)\cdot i[{\bf q}\times{\bf k}]
[{\bfg \tau}_i\times {\bfg \tau}_j]_z/m^2.
\end{eqnarray}
In the calculations, we used TVPC $NN$ amplitudes $T_{ij}$ normalized in the same way as $T$-even $P$-even amplitudes~(\ref{mpn}) and related to the amplitudes~(\ref{tpn}) as~\cite{Temerbayev:2015foa}:
\begin{equation}
\label{tt}
T_{ij} = \frac{m}{4\sqrt{\pi}k_{NN}}t_{ij},
\end{equation}
%%MP
where $k_{NN}$ is the nucleon momentum in the $NN$ center-of-mass frame and $m$ is the nucleon
mass.
Taking into account TVPC interactions, the products of $NN$ amplitudes included in the operators of normal and abnormal double scattering~(\ref{og2}), take the form
\begin{eqnarray}
\label{mmt}
[M_{ij}({\bf q}) + T_{ij}({\bf q})][M_{kl}(-{\bf q}) + T_{kl}(-{\bf q})] = \nonumber \\
 M_{ij}({\bf q}) M_{kl}(-{\bf q}) + T_{ij}({\bf q})T_{kl}(-{\bf q}) \nonumber \\
 + T_{ij}({\bf q})M_{kl}(-{\bf q}) + M_{ij}({\bf q}) T_{kl}(-{\bf q}),
\end{eqnarray}
 where the first two terms correspond to the spin-dependent $T$-even $P$-even amplitude of $dd$ scattering (the second term can actually be neglected), and the last two -- to the $T$-odd $P$-even (TVPC) amplitude.

Let's consider separately the contributions of three types of TVPC $NN$ interactions.

i) The $NN$ amplitude of the $g'$ type contributes only to the charge exchange process $pn \to np$. In $dd$ collisions, a double scattering process is possible with two sequential (or simultaneous in the case of abnormal scattering) charge-exchange collisions: $pn \to np$ and $np\to pn$. The product of the corresponding amplitudes has a form similar to Eq.~(\ref{mmt}), where $NN$ amplitudes are charge-exchange ones.
In this case, $T$-even amplitudes are the same for the processes $pn \to np$ and $np \to pn$, while $T$-odd amplitudes have an equal magnitude but an opposite sign for these two processes. Therefore, the net contribution of the $g'$-type amplitude to the TVPC signal goes to zero, as in the case of $pd$
scattering~\cite{Uzikov:2015aua}.

ii) To find the contribution of the $g$-type $NN$ amplitude, we note that due to the isospin factor $(\bm{\tau}_i - \bm{\tau}_j)_z$ (see Eq.~(\ref{tpn})), it turns to zero for identical nucleons. In this case, the operators of the normal (or abnormal) double scattering, taking into account the decomposition~(\ref{mmt}), contain the sum of $g$-type $pn$ and $np$ amplitudes multiplied by the same $T$-even $pp$ (or $pn$) amplitude. Since, due to the same isospin factor, the $g$-type amplitudes for $pn$ and $np$ elastic scattering have different signs, the net $g$-type contribution to the TVPC signal goes to zero as well. This can be easily shown by explicitly writing the operators $\hat O^{(2n)}$, $\hat O'^{(2n)}$, $\hat O^{(2a)}$ and $\hat O'^{(2a)}$ and employing the symmetry of the deuteron wave functions with respect to index permutations $1 \leftrightarrow 2$ and $3 \leftrightarrow 4$.

 iii) Thus, among the three types of TVPC $NN$ interactions, only the $h$-type amplitude contributes
 to the TVPC signal in $dd$ scattering. To calculate the respective contribution, we substitute the expansion~(\ref{mmt}) with the $h$-type TVPC $NN$ amplitude into the
 operators (\ref{og2}) and then perform integration by the nucleon coordinates
%%MP  of the nucleons of colliding deuterons -OK
in the expressions
for double-scattering amplitudes (\ref{mg2}).
%%MP
It is a straightforward but rather cumbersome procedure in the case of spin $NN$ amplitudes and the $D$-wave included in the deuterons' wave functions.
Finally, by calculating the spin matrix elements (\ref{g12m}),
we find the TVPC amplitudes of $dd$ scattering $g_i$
($i=1,2$). The detailed derivation of the $h$-type TVPC signal is given in Appendix.

As a result, we get the following expressions for the amplitudes $g_1$ and $g_2$:
\begin{eqnarray}
\label{g12}
g_1 &=& \frac{i}{2\pi m} \!\!\int\limits_0^{\infty}\!\! dq q^2 \big[Z_0 + Z(q)\big]\zeta(q) h_N(q)\big(C_n(q) + C_p(q)\big), \nonumber \\
g_2 &=& \frac{i}{2\pi m} \!\!\int\limits_0^{\infty}\!\! dq q^2 \big[Z_0 + Z(q)\big]\zeta(q) h_N(q)\big(C'_n(q) + C'_p(q)\big),
\end{eqnarray}
where the first term in square brackets refers to the normal, and the second to the abnormal double scattering.
In Eq. (\ref{g12}), we assumed $h_p = h_n = h_N$, that is justified in the beginning of the next section.
The quantities $Z_0$, $Z(q)$ and $\zeta(q)$ in Eq. (\ref{g12}) are the linear combinations of the deuteron form factors:
\begin{eqnarray}
\label{zz}
Z_0 &=& S_0^{(0)}(0) - \frac{1}{2}S_0^{(2)}(0) = 1 - \frac{3}{2}P_D, \nonumber \\
Z(q) &=& S_0^{(0)}(q) - \frac{1}{2}S_0^{(2)}(q) \nonumber \\
&&- \frac{1}{\sqrt{2}}S_2^{(1)}(q) + \sqrt{2}S_2^{(2)}(q), \nonumber \\
\zeta(q) &=& S_0^{(0)}(q) + \frac{1}{10}S_0^{(2)}(q) \nonumber \\
&& - \frac{1}{\sqrt{2}}S_2^{(1)}(q) + \frac{\sqrt{2}}{7}S_2^{(2)}(q) + \frac{18}{35}S_4^{(2)}(q),
\end{eqnarray}
where $P_D$ is the $D$-state probability in the deuteron. Note also that $Z_0 = Z(0)$.
If the $D$-wave contribution is neglected, both $Z(q)$ and $\zeta(q)$ are reduced to a purely $S$-wave form factor $S_0^{(0)}(q)$, and $Z_0$ turns to unity.
The deuteron form factors arising in (\ref{zz}) are defined as follows:
\begin{eqnarray}
\label{s02}
S_0^{(0)}(q) &=& \int\limits_0^{\infty} dr u^2(r) j_0(qr), \nonumber \\
S_0^{(2)}(q) &=& \int\limits_0^{\infty} dr w^2(r) j_0(qr), \nonumber \\
S_2^{(1)}(q) &=& 2\int\limits_0^{\infty} dr u(r)w(r) j_2(qr), \nonumber \\
S_2^{(2)}(q) &=& -\frac{1}{\sqrt{2}}\int\limits_0^{\infty} dr w^2(r) j_2(qr), \nonumber \\
S_4^{(2)}(q) &=& \frac{1}{2}\int\limits_0^{\infty} dr w^2(r) j_4(qr).
\end{eqnarray}
Note that the form factor $S_4^{(2)}(q)$ is absent in the electromagnetic structure of the deuteron.

The TVPC signal is eventually found from the
amplitudes $g_1$ and $g_2$ using the formulas~(\ref{sigtv}) for a given combination of polarizations of colliding deuterons.

\section{\label{sec3} Numerical results}

For numerical calculations, spin amplitudes
of $pp$ and $pn$ elastic scattering are required, both $T$-even $P$-even amplitudes from (\ref{mpn}) and
$T$-odd $P$-even ones from (\ref{tpn}).
As was shown in the previous section, the interactions of $g$
and $g'$ type do not give a contribution to the TVPC signal in $dd$ scattering. Therefore, we consider here only the $h$-type interaction.
The numerical value of the constant in the
respective
amplitude $h_N$ (\ref{tpn}) is unknown, so it is impossible to calculate the absolute value of the TVPC signal, but it is possible to calculate
its dependence on
the collision energy.

The amplitude $h_N$ dependence
on the momentum $q$ can be given under assumption that the $h$-type $NN$ interaction is determined by the exchange of  the $h_1(1170)$ meson with quantum numbers
 $I^{G}(J^{PC})=0^-(1^{+-})$
between nucleons
(see \cite{Uzikov:2015aua} and references therein).
 % Simonius 1975
Under this assumption,
according to
the works \cite{Uzikov:2015aua, Beyer93},
we take the following expression for the amplitude $h_{N}$:
 \begin{equation}
  h_N = -i\phi_h\frac{2G_h^2}{m_h^2+{\bf q}^2} F_{hNN}({\bf q}^2),
 \end{equation}
where  $\phi_h=\tilde G_h/G_h$ is
the ratio of the coupling constant of the $h_1$-meson
with a nucleon for the $T$-non-invariant interaction ($\tilde G_h$) to the corresponding constant of the $T$-invariant interaction
($G_h$);
$F_{hNN}({\bf q}^2)=(\Lambda^2-m_h^2)/(\Lambda^2-{\bf q}^2)$ is
the phenomenological monopole form factor at the $hNN$ vertex. The numerical parameters are taken from
Ref.~\cite{Beyer93}:
$m_h=1.17$ GeV,
 $G_h=4\pi\times 1.56$,
 { and  $\Lambda=2$} GeV, as follows from the Bonn $NN$-interaction potential.
 At the same time, due to isoscalar nature of this meson, we have the equality of the amplitudes $h_p=h_n=h_N$,
 which is taken into account
 in the formulas (\ref{g12}) for the $dd$ TVPC amplitudes $g_i$ ($i=1,2$).

In the range of laboratory proton beam energies
0.1--1.2 GeV in $pN$ scattering
(corresponding to the interval of the invariant mass of colliding nucleons $\sqrt{s_{pN}}=1.9$--$2.4$ GeV), the $T$-even $P$-even amplitudes $A_N,
   \cdots, H_N$ are available in the SAID database \cite{Arndt:2007qn}, which we use here in the numerical calculations of the TVPC signal
   at these energies.
In the calculations at higher energies $\sqrt{s_{NN}} \gtrsim 2.5$ GeV corresponding to the conditions of the NICA SPD experiment,
we employ the phenomenological models for the spin amplitudes of $pN$ elastic scattering available in the literature.

In the formulation of $pp$ scattering models in the high-energy region, the helicity amplitudes $\phi_1\div \phi_5$ are used, in the
conventional notation (see
  \cite{Sibirtsev:2010cjv}).
  The spin amplitudes $A_N$, $B_N$, $C_N$, $C'_N$, $G_N$ and $H_N$, defined in (\ref{mpn}),
are related to the helicity amplitudes by the following relations
valid at small momentum transfers and high energies specific for the Glauber model (see
\cite{Platonova:2010wjt} and references therein):
\begin{eqnarray}
A_N=(\phi_1+\phi_3)/2,\nonumber \\
B_N=(\phi_3-\phi_1)/2, \nonumber \\
C_N=i\phi_5, \ \
G_N=\phi_2/2, \ \ H_N=\phi_4/2; \nonumber \\
C'_N=C_N+\frac{iq}{2m}A_N.
\label{A-phi}
\end{eqnarray}

Here we use two
different models for the helicity amplitudes of $pN$ elastic scattering.
The first one
\cite{Sibirtsev:2010cjv} involves the Regge
parametrization of data on the $pp$ differential cross section and spin correlations $A_N$, $A_{NN}$
in the range of laboratory momenta $3\div 50$ GeV/$c$.
  This model includes the contributions from four Regge trajectories, $\omega$, $\rho$, $f_2$, $a_2$, and the $P$ pomeron exchange.
   As noted in \cite{Sibirtsev:2010cjv}, in the Regge model, because of isospin symmetry and relations due to $G$ parity, $pp$ and $pn$ scattering amplitudes can be represented as the following linear
   combinations of these five contributions:
 \begin{eqnarray}
 \label{pn}
  \phi(pp)= -\phi_\omega -\phi_\rho + \phi_{f_2} +\phi_{a_2} +\phi_P,\nonumber \\
   \phi(pn)= -\phi_\omega +\phi_\rho + \phi_{f_2} -\phi_{a_2} +\phi_P;
   \end{eqnarray}
here, for example, $\phi_\omega$ is the contribution of the $\omega$ Regge trajectory, etc.
   The energy domain, in which the Regge parametrization was performed in \cite{Sibirtsev:2010cjv},
  corresponds to
  the range of the $pp$ invariant mass
$\sqrt{s_{pp}}= 2.8$--$10$ GeV.

   The second approach used here is based on
 the Regge-eikonal model developed by O. Selyugin
(see the work \cite{Selyugin:2021cem} and references therein) and is called by its author
the HEGS (High Energy Generalized Structure) model. This model considers $pp$, $p\bar{p}$ and $pn$ elastic scattering at small angles taking into account the nucleon structure based on the
data on generalized parton distributions of nucleons.
The helicity amplitudes of $NN$ elastic scattering obtained in this model allow one to describe the available
   experimental data on the differential cross section and single-spin asymmetry $A_N(s,t)$ in $pp$ scattering
    in the energy range $\sqrt{s}$ from 3.6 to 10 TeV with a minimum of variable parameters
   \cite{Selyugin:2024ccc}
In both models, at the energies
$\sqrt{s_{pp}}\ge 3$ GeV considered here, the following
approximate relations hold for the helicity amplitudes of $pp$ elastic scattering:
$\phi_1=\phi_3$, $\phi_2=0,\phi_4=0$.

 When calculating the TVPC signal according to the optical theorem, the Coulomb contributions are excluded here from $pp$ amplitudes. The explanation for this is given in
 \cite{Uzikov:2015aua, Song:2016wwa}.
The reason is that the Coulomb
interaction does not violate the $T$ invariance and therefore cannot make a direct contribution to the TVPC signal.
Indeed, the spin structure of the transition operator for
scattering on the deuteron at zero angle is such that
the spin-independent amplitude $A_N$ and the amplitudes $B_N$, $G_N$, $H_N$, additively containing the Coulomb
contribution,
do not enter
the expressions for $dd$ TVPC amplitudes (\ref{g12}).
At the same time,
the Coulomb term
enters the spin-flip amplitude $C'_N$ through the amplitude $A_N$, however, $A_N$ is multiplied here by the transferred momentum $q$ (see Eq. (\ref{A-phi})), which compensates for the Coulomb singularity at $q\to 0$ when integrating over $q$ in Eq. (\ref{g12}).  Numerically, the contribution of the Coulomb interaction to the TVPC signal is negligible
\cite{Uzikov:2015aua}.

\begin{figure}[!ht]
\begin{center}
\includegraphics[width=127mm]{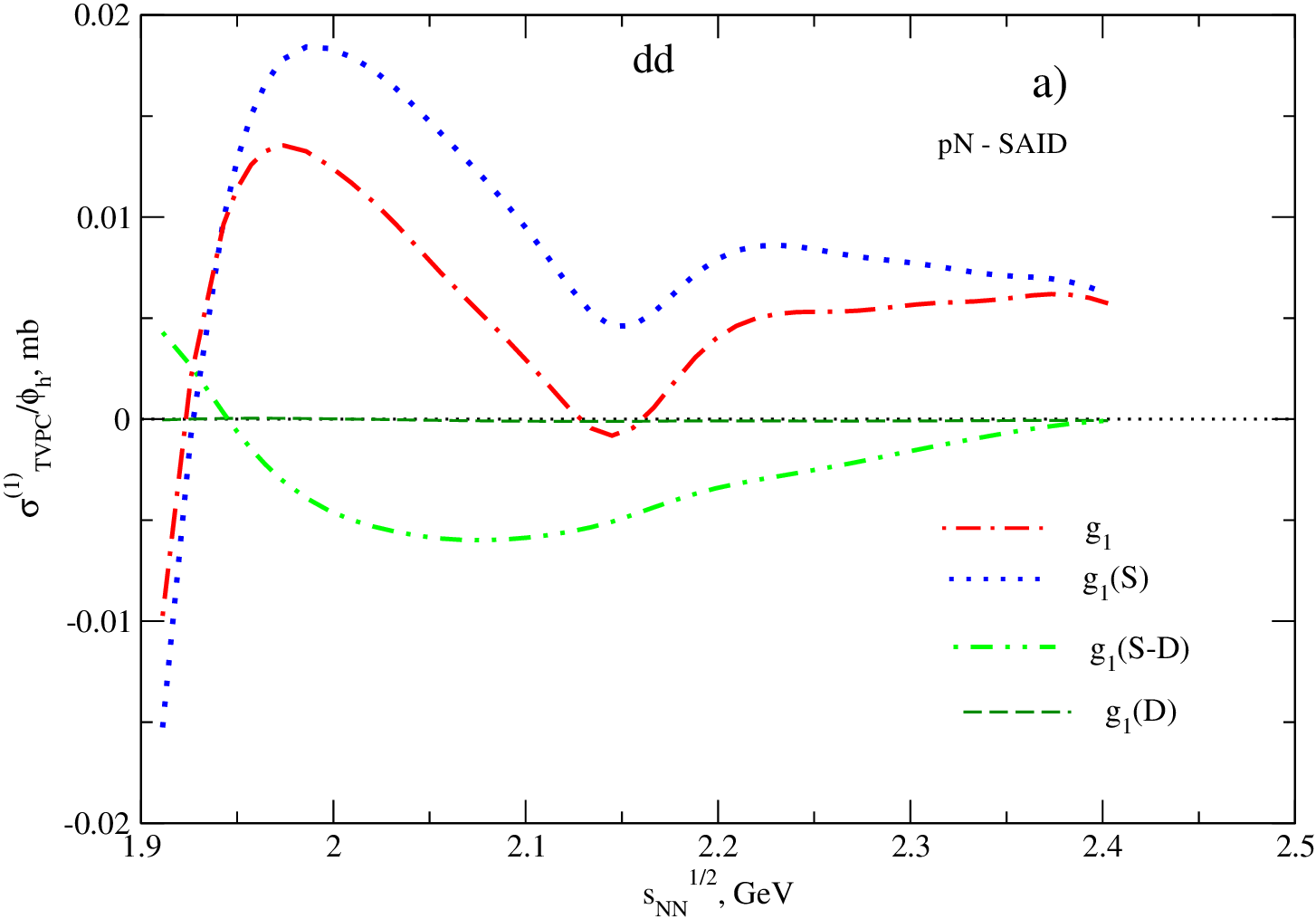}
\includegraphics[width=127mm]{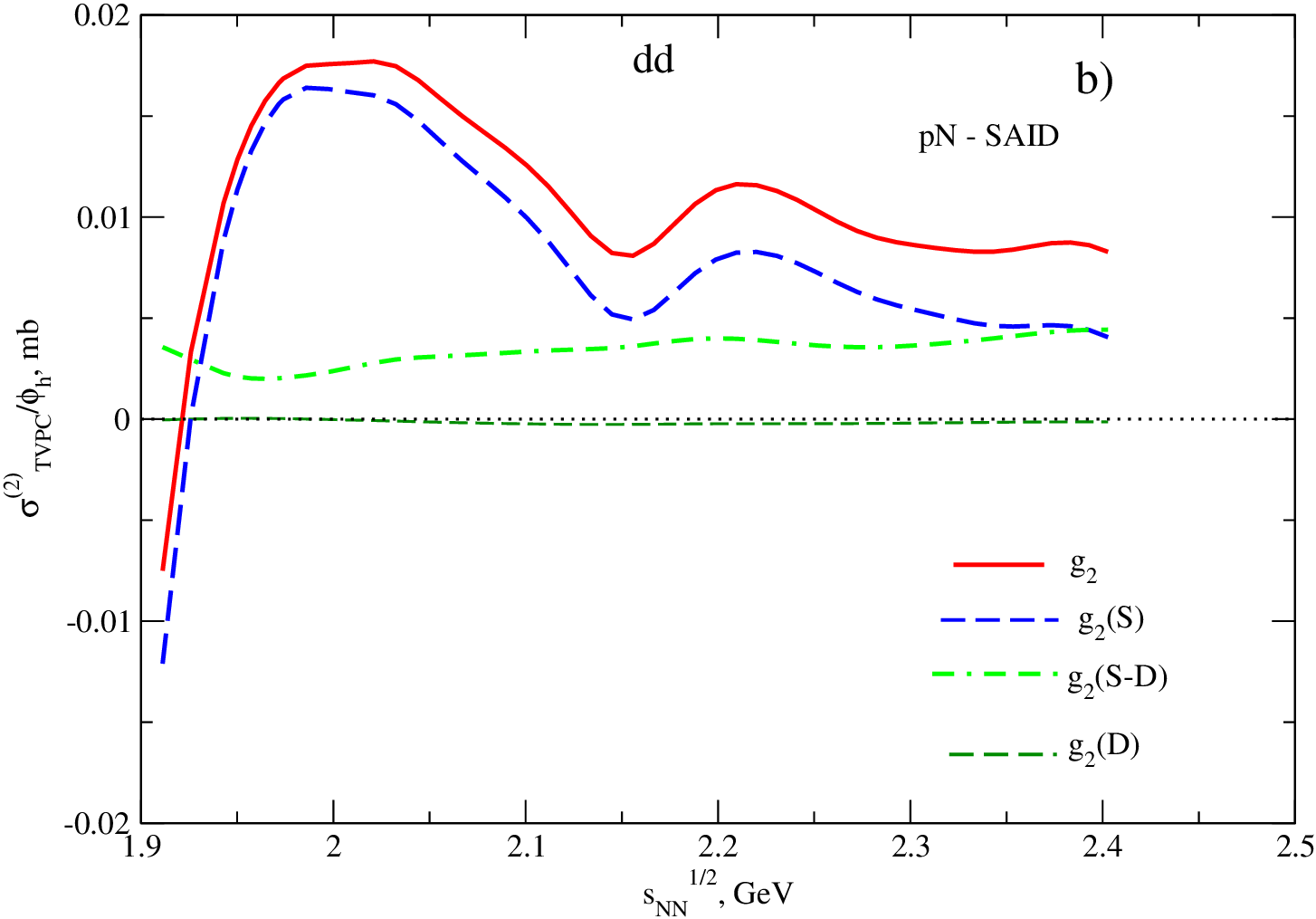}
\vspace{-3mm}
\caption{(color online)
Energy dependence of TVPC signals
(cross sections) corresponding to the amplitudes
$g_1$ (a) and $g_2$ (b) in $dd$ scattering for spin $pN$ amplitudes taken from the SAID database \cite{Arndt:2007qn}.
(a) $g_1$: $S$-wave (dotted line),
$D$-wave (thin dashed), $S$-$D$ interference (dash-dot-dotted),
total $S+D$ (dash-dotted);
(b) $g_2$: $S$-wave (dashed line),
$D$-wave (thin dashed), $S$-$D$ interference (dash-dash-dotted),
total $S+D$ (solid).
The invariant mass of the interacting $NN$ pair -- one nucleon from the beam and
another one
from the target -- is shown along the $X$-axis.
On both panels, a straight thin dotted line shows the zero level for easy visualization.
\label{fig01}
}
\end{center}
\vspace{-5mm}
\end{figure}

\begin{figure}[!ht]
\begin{center}
%%\resizebox{0.6\columnwidth}{!}
%%%\includegraphics[width=127mm]{dd-Sib-full.eps
%\includegraphics[width=127mm]{dd-Sib-full.eps}
%
\includegraphics[width=127mm]{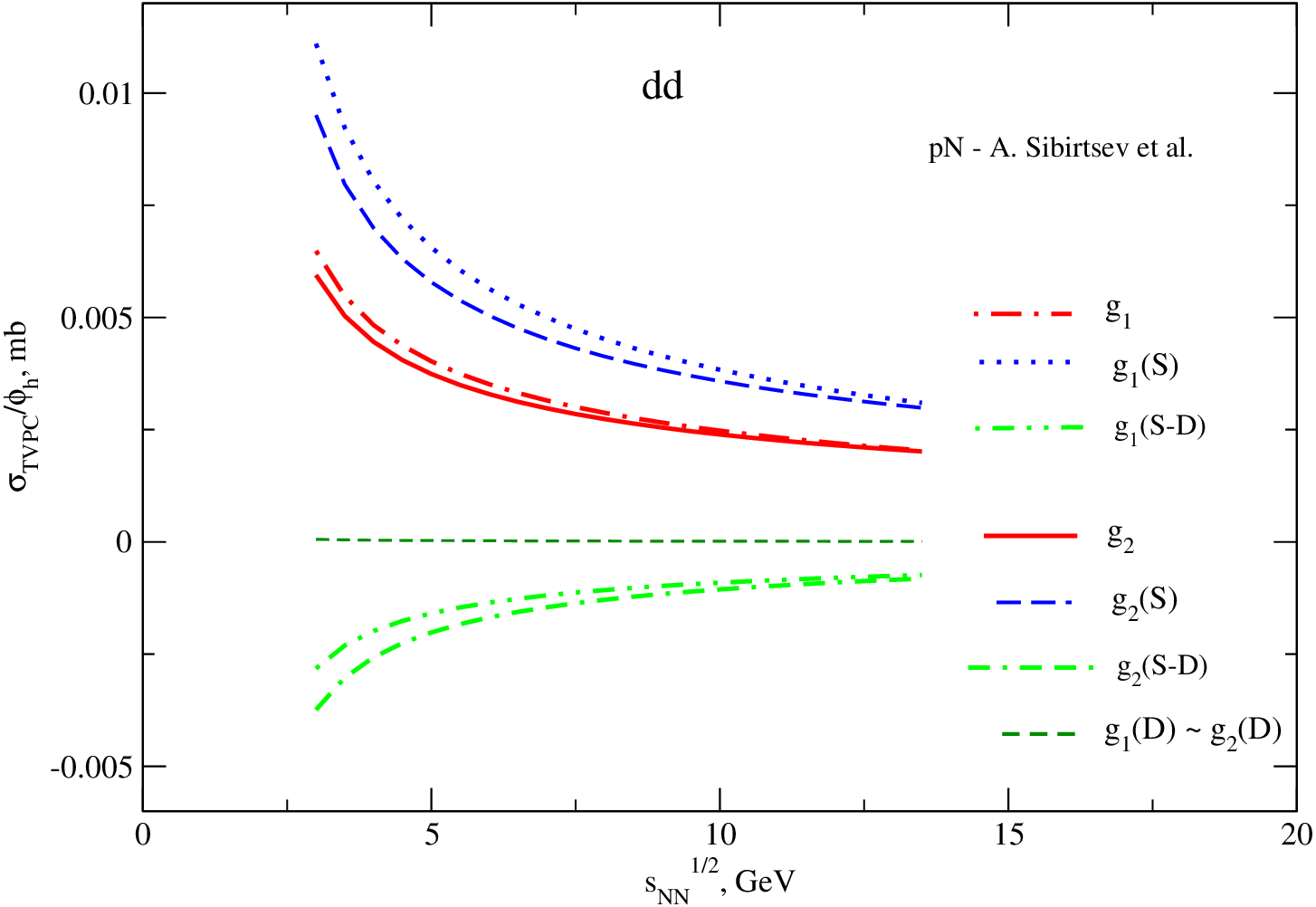}
\vspace{-3mm}
\caption{
(color online)
Energy dependence of TVPC signals corresponding to $g_1$ and $g_2$ amplitudes in $dd$ scattering for spin $pN$ amplitudes taken from
Ref. \cite{Sibirtsev:2010cjv}. Notations are the same as in Fig.~\ref{fig01}, panels (a) and (b). \label{fig02}
}
\end{center}
\vspace{-5mm}
\end{figure}

\begin{figure}[!ht]
\begin{center}
\includegraphics[width=127mm]{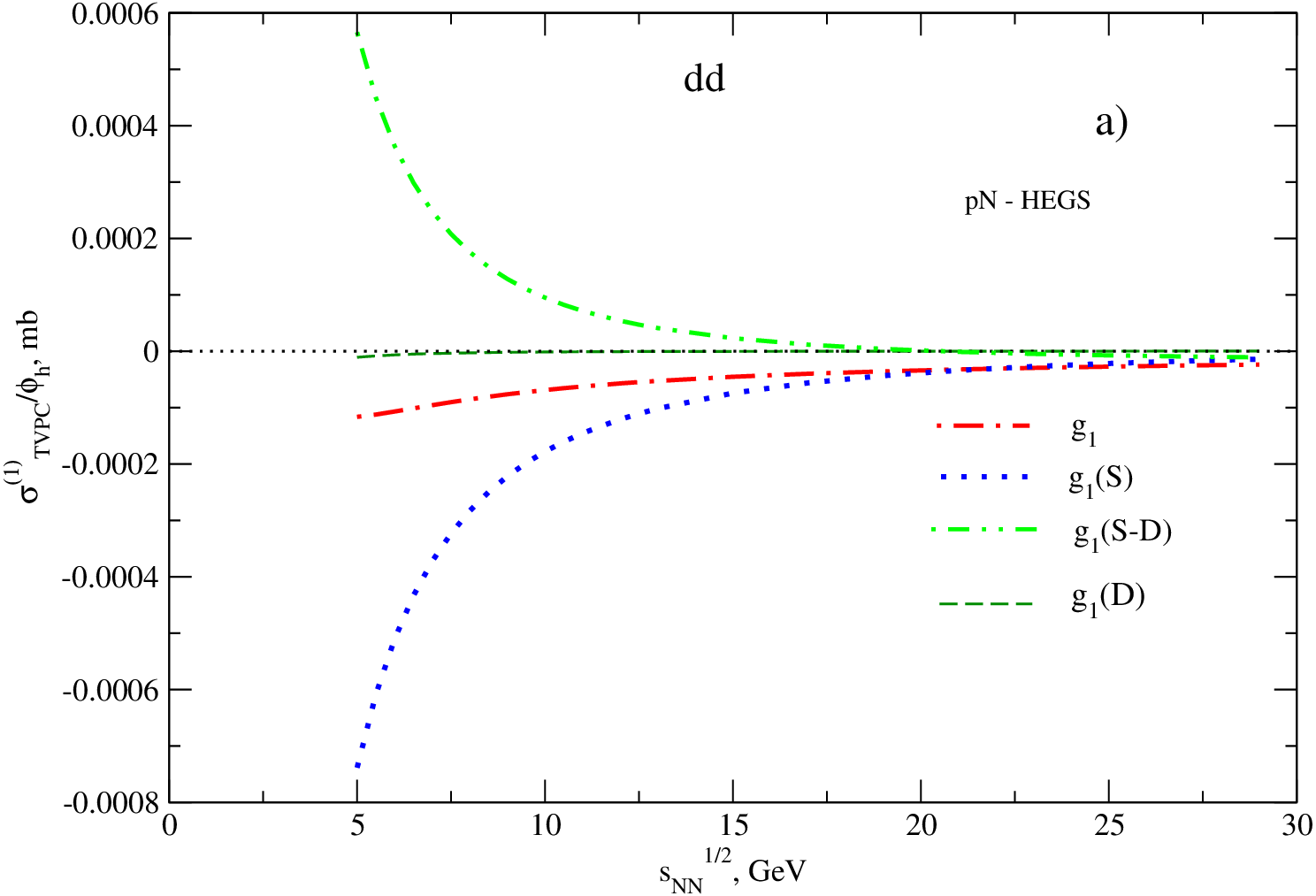}
\includegraphics[width=127mm]{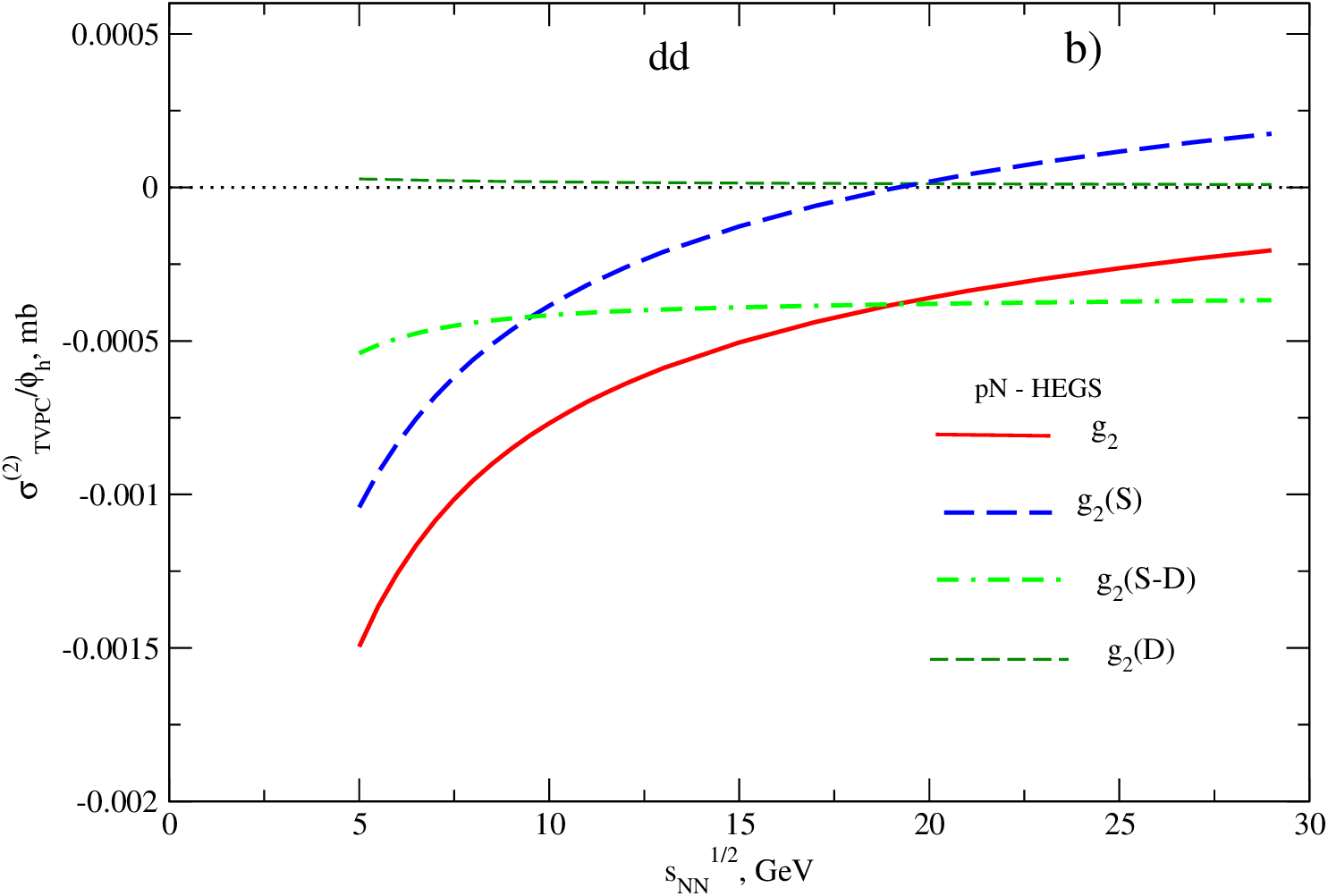}
\vspace{-3mm}
\caption{(color online)
The same as in Fig.~\ref{fig01} but for spin $pN$ amplitudes
from the HEGS model
%\cite{Selyugin:2021cem}
\cite{Selyugin:2024ccc}. \label{fig03}
}
\end{center}
\vspace{-5mm}
\end{figure}

The figures show the results of our calculations of the TVPC signal in $dd$ scattering using the SAID database
(Fig. \ref{fig01}) and two
phenomenological
models for spin $NN$ amplitudes -- Regge parametrization (Fig. \ref{fig02}) and HEGS model (Fig. \ref{fig03}) in the energy intervals corresponding to these parametrizations.

It is seen from Fig. \ref{fig01} that
the maximum of the signal is located in the energy range of
1.95--2.05 GeV and its absolute value
%%YU
is unevenly decreasing with further increase of collision energy
 although demonstrating the second local maximum at $\sim 2.2$ GeV in $\sigma^{(2)}_{TVPC}$
and some a plateau in $\sigma^{(1)}_{TVPC}$.

%%MP Здесь убывание неравномерное, для g_1 оно вообще неочевидно из рисунка при больших энергиях. Стоит ли написать о локальном минимуме при 2.15 ГэВ или об убывании сигнала до этой энергии?
%%YU - Да,НАдо отметить  минимум  и новый локальный  максимси или %плато.
One should note
that the $S$-wave of the deuteron dominates
in both amplitudes $g_1$ and $g_2$ in the entire
range of the invariant mass
$\sqrt{s_{NN}}= 1.9$--$2.4$ GeV
covered by the SAID database,
whereas
the contribution of the pure $D$-wave is negligible.
%%MP and the $S-D$ interference is destructive  - только для g_1!
The $S-D$ interference is essential and
destructive for the $g_1$ amplitude, but constructive for the $g_2$ one.
%%MP - OK. Я писала это для себя и уже проверила и исправила (в абзаце ниже была путаница в дейтронах 1,2).
%% Сейчас вставила еще в скобках ($g_i$), чтобы было совсем понятно, где какая амплитуда.
%% Можно вставить еще ссылку на формулу (2).
%%:
  The numerical difference between the amplitudes $g_1$ and $g_2$ is due to the fact that one of them
%%MP
  ($g_2$)
  is calculated in the rest
  frame of a tensor polarized ($P_{xz}^{(2)}$) deuteron target $d_2$,
on which a vector polarized ($P_y^{(1)}$)
deuteron beam
$d_1$
scatters,
and the other ($g_1$) -- in a collision, when
a tensor-polarized ($P_{xz}^{(1)}$) deuteron beam $d_1$ falls on a vector-polarized ($P_y^{(2)}$) target $d_2$.

The results obtained by
using the Regge parametrization of $pN$ amplitudes from the work \cite{Sibirtsev:2010cjv}
are shown in Fig. \ref{fig02}. With this $pN$ input,
 the amplitudes $g_1$ and $g_2$ are numerically very close to each other
 separately for the $S$- and $D$-wave
contributions, as well as for the total $S+D$ calculation.
The $D$ wave is negligible and the $S-D$ interference is destructive for both $g_1$ and $g_2$ amplitudes.
 It should be noted that the maximum of the TVPC signal
 is obtained at the minimal energy $\sqrt{s_{NN}}=2.6$ GeV from the range considered,
and the signal decreases
%%MP rapidly
monotonically with an increase of the collision energy $\sqrt{s_{NN}}$.

 With the HEGS parametrization
 \cite{Selyugin:2021cem,Selyugin:2024ccc}, at energies
 $\sqrt{s_{NN}} \sim 5$ GeV
the TVPC signal is obtained about an order of magnitude less than with the parametrization \cite{Sibirtsev:2010cjv} and also decreases with increasing energy
(see Fig.~\ref{fig03}).
Like for the parametrization
from
Refs. \cite{Arndt:2007qn} and \cite{Sibirtsev:2010cjv},
when using the HEGS model, the contribution of
the deuteron $D$ wave to the TVPC signal is negligible in magnitude
in comparison to the $S$-wave contribution, and $S-D$ interference is significant. Furthermore,
as for the parametrization from
Ref. \cite{Arndt:2007qn},
the $S-D$ interference is
destructive for the $g_1$ amplitude and constructive for the $g_2$ one.

%\begin{figure}[!ht]
%\begin{center}
%\resizebox{0.6\columnwidth}{!}
%{\includegraphics[width=1.6in]{dd-Saidm-full.eps}}
%\resizebox{0.67\columnwidth}{!}{\includegraphics{fig1-a.eps}}
%\vfill
%\vspace{0.6cm}
%\resizebox{0.8\columnwidth}{!}{\includegraphics{p3He-200-dsig.eps}}
%\vfill
%\resizebox{0.67\columnwidth}{!}{\includegraphics{fig1-b.eps}}
%\vfill
%\vspace{0.6cm}
%\resizebox{0.8\columnwidth}{!}{\includegraphics{p3He-415-dsig.eps}}
%\vfill
%\resizebox{0.67\columnwidth}{!}{\includegraphics{fig1-cm.eps}}
%\vfill
%\vspace{0.6cm}
%\resizebox{0.6\columnwidth}{!}{\includegraphics{p3He-1000-dsig.eps}}
%\resizebox{0.67\columnwidth}{!}{\includegraphics{fig1-d.eps}}
%\end{center}
%\caption{Дифференциальные сечения упругого $p{}^3$He-рассеяния при энергиях
%\end{figure}

\section{\label{sec4} Conclusion}

In this paper, the signal of $T$-invariance violation {
%with $P$-invariance  conserving}
with {parity} conserved}
(TVPC) has been calculated (up to an unknown constant) for $dd$ scattering.
The calculation is based on the Glauber diffraction theory with full consideration of the spin dependence of the $NN$ scattering amplitudes.
We take into account the contributions of single and double scattering mechanisms dominating in the amplitude of the elastic process $dd\to dd$ in the region of the first diffraction maximum, which gives the main contribution to the TVPC signal~\cite{Uzikov:2024}.
For the first time, the $D$-component of the deuteron wave function is taken into account in the calculation of this effect together with the $S$-component accounted for earlier in~\cite{Uzikov:2024}. The $S$-$D$ interference is found to be significant in the TVPC signal.

The TVPC scattering amplitude is much
smaller in magnitude as compared to the corresponding $T$-even hadron amplitude.
However, due to different symmetry properties of these amplitudes,
the $T$-odd amplitude of
elastic scattering
does not interfere with the corresponding $T$-even one.
Therefore, { the typical accuracy of a Glauber theory
   calculation of the total cross section
is similar to that of the calculation of the TVPC signal}.
To a large extent, this accuracy
is determined by our knowledge of $NN$ elastic scattering amplitudes, which are included in the TVPC signal as multipliers.

Here, for the $pN$ amplitudes,
we used the database \cite{Arndt:2007qn}
at lower energies and an available
parametrization \cite{Sibirtsev:2010cjv} and a phenomenological model \cite{Selyugin:2024ccc} at higher energies.
 The energy ranges of $pN$ collisions correspond to the intervals of the invariant mass of the $NN$ pair
$\sqrt{s_{NN}}= 1.9$--$2.4$ GeV (the laboratory kinetic energy of the proton $T_l=0.1$--$1.2$ GeV) and
$\sqrt{s_{NN}}= 2.5$--$25$ GeV (the laboratory momentum of the proton beam $P_{l}=2.2$--$332$ GeV/$c$.)

It has been shown that the maximum value of the TVPC signal
corresponds to the invariant mass $\sqrt{s_{NN}} \sim 1.95$--$2.05$ GeV. At { the} collision energies corresponding to the conditions of the SPD NICA experiment, $\sqrt{s_{NN}} \gtrsim 2.5$ GeV, the magnitude of the signal depends essentially on the model used for the $T$-even $P$-even spin amplitudes of $pN$ scattering and decreases with increasing energy, under an assumption that the TVPC interaction constant does not depend on energy. This
is consistent with the general trend of spin phenomena -- the decrease of the
$T$-even $P$-even
spin effect in magnitude with increasing energy. However, at the energies of the NICA complex corresponding to the conditions of the early baryon Universe, a possible growth of an unknown TVPC constant is not excluded.

We have found that { only one of the three types of the TVPC $NN$ interaction that do not disappear on the mass shell, i.e., $h_N$, gives a non-zero contribution to the TVPC signal}, while the contributions of other two
($g_N$ and $g'_N$) vanish due to the specific symmetry properties of these interactions.
By this property, the search for a TVPC signal in
$dd$ scattering differs
from the previously considered processes of $pd$ and
$^3$He$d$ scattering, where two types of { the} TVPC $NN$ interaction, $h_N$ and $g_N$, give the non-zero contributions
\cite{Uzikov:2015aua, Uzikov:2016lsc, Uzikov:2023eex}.
This is one of the main results of this work, which is important for extracting the unknown constant of the TVPC interaction from the respective data.

\begin{acknowledgments}
The authors are grateful to O.V. Selyugin for providing the files with numerical values of spin $pN$ amplitudes obtained in the model developed by him.
\end{acknowledgments}

\appendix*
\section{Derivation for $h$-type TVPC amplitudes}
To find the operators of normal and abnormal double scattering~(\ref{og2}) in the case of TVPC $NN$ interaction of the $h$ type, we use the expression~(\ref{mmt}) and omit the terms linear in $\hat{\bf q}$ (or $\hat{\bf n}$), which turn to zero when integrated over the direction of the vector ${\bf q}$ in~(\ref{mg2}).
Then, taking into account the symmetry of the deuteron wave functions in respect of permutation of the nucleon indices, the spin dependence of the operators~(\ref{og2}) can be represented as (here and forth, under the $\hat O^{(2n)}$, $\hat O'^{(2n)}$, etc., we mean operators for the $h$-type TVPC interaction):
\begin{eqnarray}
\label{og2v}
\hat O^{(2n)}({\bf q}) &=& \bm{\sigma}_1\cdot {\bf V}_n(\bm{\sigma}_3,\bm{\sigma}_4), \nonumber \\
\hat O'^{(2n)}({\bf q}) &=& \bm{\sigma}_3\cdot {\bf V}'_n(\bm{\sigma}_1,\bm{\sigma}_2), \nonumber \\
\hat O^{(2a)}({\bf q}) &=& \bm{\sigma}_1\cdot {\bf V}_a(\bm{\sigma}_3,\bm{\sigma}_4), \nonumber \\
\hat O'^{(2a)}({\bf q}) &=& \bm{\sigma}_3\cdot {\bf V}'_a(\bm{\sigma}_1,\bm{\sigma}_2),
\end{eqnarray}
where
\begin{eqnarray}
\label{vvp}
{\bf V}_n(\bm{\sigma}_3,\bm{\sigma}_4) \!\!&=&\!\! -2\Pi (h_p C_n + h_n C_p) \nonumber \\
&&\!\!
\times[\hat{\bf k}(\bm{\sigma}_3\!\cdot\!\hat{\bf q})(\bm{\sigma}_4\!\cdot\!\hat{\bf n}) + \hat{\bf q}(\bm{\sigma}_3\!\cdot\!\hat{\bf k})(\bm{\sigma}_4\!\cdot\!\hat{\bf n})], \nonumber \\
{\bf V}'_n(\bm{\sigma}_1,\bm{\sigma}_2) \!\!&=&\!\! -2\Pi (h_p C'_n + h_n C'_p) \nonumber \\
&&\!\!\times[\hat{\bf k}(\bm{\sigma}_1\!\cdot\!\hat{\bf q})(\bm{\sigma}_2\!\cdot\!\hat{\bf n}) + \hat{\bf q}(\bm{\sigma}_1\!\cdot\!\hat{\bf k})(\bm{\sigma}_2\!\cdot\!\hat{\bf n})]
\end{eqnarray}
and $\Pi = \frac{q}{4\sqrt{\pi} m}$.
The vector operators ${\bf V}_a(\bm{\sigma}_3,\bm{\sigma}_4)$ and ${\bf V}'_a(\bm{\sigma}_1,\bm{\sigma}_2)$
are similar to
${\bf V}_n(\bm{\sigma}_3,\bm{\sigma}_4)$ and ${\bf V}'_n(\bm{\sigma}_1,\bm{\sigma}_2)$, respectively,
with the replacement
$h_n \leftrightarrow h_p$.

Such a representation allows one to easily integrate over the coordinates of nucleons inside one of the colliding deuterons.
Thus, after integrating the
normal double-scattering operator $\hat O^{(2n)}$ over the coordinates of nucleons in the target, we obtain the operator
\begin{eqnarray}
\label{onrh}
\hat\Omega^{(2n)}({\bf q}) &=& \int d^3 r \Psi^+_{d(12)}({\bf r}) \hat O^{(2n)}({\bf q}) \Psi_{d(12)}({\bf r}) \nonumber \\
 &=& Z_0 {\bf S}^{(2)}\cdot{\bf V}_n(\bm{\sigma}_3,\bm{\sigma}_4),
\end{eqnarray}
where ${\bf S}^{(2)}$ is
the spin operator of the target deuteron and the factor $Z_0$ is defined in (\ref{zz}).
For the $\hat O'^{(2 n)}$, we obtain a similar expression after integration by the coordinates of nucleons in the beam:
\begin{eqnarray}
\label{onr}
\hat\Omega'^{(2n)}({\bf q}) &=& \int d^3 \rho \Psi^+_{d(34)}(\bm{\rho}) \hat O'^{(2n)}({\bf q}) \Psi_{d(34)}(\bm{\rho}) \nonumber \\
 &=& Z_0 {\bf S}^{(1)}\cdot{\bf V'}_n(\bm{\sigma}_1,\bm{\sigma}_2),
\end{eqnarray}
where  ${\bf S}^{(1)}$  is
the spin operator of the incident deuteron.

In the same way, the abnormal double-scattering operator $\hat O^{(2a)}$ is integrated by $d^3r$ (with a factor of $e^{i\bf{qr}}$), and $\hat O'^{(2a)}$ -- by $d^3\rho$ (with the factor $e^{-i\bf{q}\bm{\rho}}$), and we get the following expressions:
\begin{eqnarray}
\label{oarh0}
\hat\Omega^{(2a)}({\bf q}) &=& \int d^3 r \Psi^+_{d(12)}({\bf r}) e^{i\bf{qr}} \hat O^{(2a)}({\bf q}) \Psi_{d(12)}({\bf r}) \nonumber \\
&=& [S_0^{(0)}(q) - \frac{1}{2}S_0^{(2)}(q)]{\bf S}^{(2)}\cdot{\bf V}_a(\bm{\sigma}_3,\bm{\sigma}_4) \nonumber \\
&& + \frac{1}{\sqrt{2}}[S_2^{(2)}(q) - \frac{1}{\sqrt{2}}S_2^{(1)}(q)] \nonumber \\
&&\times \hat S_{12}(\hat{\bf q}; {\bf S}^{(2)}, {\bf V}_a(\bm{\sigma}_3,\bm{\sigma}_4)),
\end{eqnarray}
\begin{eqnarray}
\label{oar0}
\hat\Omega'^{(2a)}({\bf q}) &=& \int d^3 \rho \Psi^+_{d(34)}(\bm{\rho}) e^{-i\bf{q}\bm{\rho}} \hat O'^{(2a)}({\bf q}) \Psi_{d(34)}(\bm{\rho}) \nonumber \\
&=&[S_0^{(0)}(q) - \frac{1}{2}S_0^{(2)}(q)]{\bf S}^{(1)}\cdot{\bf V}'_a(\bm{\sigma}_1,\bm{\sigma}_2) \nonumber \\
&& + \frac{1}{\sqrt{2}}[S_2^{(2)}(q) - \frac{1}{\sqrt{2}}S_2^{(1)}(q)] \nonumber \\
&&\times \hat S_{12}(\hat{\bf q}; {\bf S}^{(1)}, {\bf V}'_a(\bm{\sigma}_1,\bm{\sigma}_2)),
\end{eqnarray}
where  the deuteron form factors $S_i^{(j)}(q)$ are defined in Eq. (\ref{s02}).

Next, we note that in a calculation of the TVPC signal, only non-diagonal spin matrix elements~(\ref{g12m}) are needed. Therefore, the components of the vectors ${\bf V}_a$ and ${\bf V}'_a$ parallel to $\hat{\bf k}$ (see the definition (\ref{vvp}) and the text below it) do not contribute to the TVPC signal (with the standard choice of $\hat{\bf k} || Oz$).
For the component ${\bf V}'_a$ parallel to $\hat{\bf q}$ (we denote it as ${\bf V}'^q_a$), one has:
\begin{equation}
\label{svaq}
\hat S_{12}(\hat{\bf q}; {\bf S}^{(1)}, {\bf V}'^q_a) = 2 {\bf S}^{(1)}\cdot{\bf V}'^q_a,
\end{equation}
%\begin{eqnarray*}
%<m_1'=-1|\hat S_{12}(\hat{\bf q}; {\bf S}^{(1)}, {\bf V}'_a)|m_1=0> = \nonumber \\
%2 <m_1'=-1|{\bf S}^{(1)}\cdot{\bf V}'_a|m_1=0>,
%\end{eqnarray*}
and a similar relation is fulfilled for the component ${\bf V}^q_a$ (with the replacement
${\bf S}^{(1)} \to {\bf S}^{(2)}$).
Let's denote the parts of the operators~(\ref{oarh0}) and~(\ref{oar0}), including only the components ${\bf V}^q_a$ and ${\bf V}'^q_a$, via $\hat\Omega^{(2a)}_q({\bf q})$ and $\hat\Omega'^{(2a)}_q({\bf q})$, respectively. Taking into account the relations~(\ref{svaq}), we get the expressions for them similar to~(\ref{onrh}) and~(\ref{onr}), respectively:
\begin{eqnarray}
\label{aarh}
\hat\Omega^{(2a)}_q({\bf q}) &=& Z(q) {\bf S}^{(2)}\cdot{\bf V}^q_a(\bm{\sigma}_3,\bm{\sigma}_4),
\end{eqnarray}
\begin{eqnarray}
\label{aar}
\hat\Omega'^{(2a)}_q({\bf q}) &=& Z(q) {\bf S}^{(1)}\cdot{\bf V}'^q_a(\bm{\sigma}_1,\bm{\sigma}_2),
\end{eqnarray}
where the factor $Z(q)$ is defined in (\ref{zz}).

We now proceed to integration by the coordinates of the nucleons inside the second deuteron.
To do this, it is convenient to represent the vector ${\bf V}'_n(\bm{\sigma}_1,\bm{\sigma}_2)$ (see Eq.~(\ref{vvp})) as
\begin{equation}
\label{vw}
{\bf V}'_n(\bm{\sigma}_1,\bm{\sigma}_2) = {\bf W}'^n_{ij}\sigma_{1i}\sigma_{2j},
\end{equation}
where
\begin{equation}
\label{wij}
{\bf W}'^n_{ij} = - 2\Pi(h_p C'_n + h_n C'_p)[\hat{\bf k} \, {\hat q}_i {\hat n}_j + \hat{\bf q} \, {\hat k}_i {\hat n}_j].
\end{equation}
Similarly, ${\bf V}_n(\bm{\sigma}_3,\bm{\sigma}_4) = {\bf W}^n_{ij}\sigma_{3i}\sigma_{4j}$,
where ${\bf W}^n_{ij}$ has the same form (\ref{wij}), but with the replacement of $C'_N \to C_N$.
In turn, for the vectors ${\bf V}_a$ and ${\bf V'}_a$, we introduce a similar representation with ${\bf W}^a_{ij}$ and ${\bf W}'^a_{i j}$, which differ from ${\bf W}^n_{ij}$
and ${\bf W}'^n_{ij}$ by replacing $h_n \leftrightarrow h_p$ only.
Such a representation allows integration by
the nucleon coordinates inside the second deuteron in the same way as it was done
for $pd$ scattering
(for example,
%%MP see - здесь должно быть именно using
 using the
formula (12) from Ref.~\cite{Platonova:2010zz}).

Thus, by integrating the operator $\hat\Omega'^{2n}({\bf q})$ (\ref{onr}) with the factor $e^{i{\bf qr}}$ over the nucleon coordinates inside the target deuteron and employing the definition of the deuteron form factors (\ref{s02}), we obtain:

\begin{widetext}
\begin{eqnarray}
\label{onq}
\int d^3 r \Psi^+_{d(12)}({\bf r}) e^{i{\bf qr}}\hat\Omega'^{2n}({\bf q})\Psi_{d(12)}({\bf r}) = Z_0\Biggl(\left[S_0^{(0)}(q) -\frac{1}{2}S_0^{(2)}(q)+ \frac{1}{2\sqrt{2}}S_2^{(1)}(q) + \frac{7}{2\sqrt{2}}S_2^{(2)}(q)\right]{\bf S}^{(1)}\!\cdot\!{\bf W}'^n_{ij}\{S^{(2)}_i,S^{(2)}_j\} \nonumber \\
+ \frac{3}{2\sqrt{2}}\left[S_2^{(1)}(q) + S_2^{(2)}(q)\right]{\bf S}^{(1)}\!\cdot\!{\bf W}'^n_{ij}\{[{\bf S}^{(2)}\times {\hat q}]_i,[{\bf S}^{(2)}\times {\hat q}]_j\}
 + 3{\bf S}^{(1)}\!\cdot\!{\bf W}'^n_{ij} \!\!\int\!\! d^3 r \frac{e^{i{\bf qr}}}{4\pi r^2}w^2(r)\hat S_{12}(\hat{\bf r}; {\bf S}^{(2)},{\bf S}^{(2)}){\hat r}_i{\hat r}_j\Biggr),
\end{eqnarray}
\end{widetext}
where it was taken into account that
${\bf W}'^n_{ij} \delta_{ij} = {\bf W}'^n_{ij} \hat q_i \hat q_j = 0$ (see Eq. (\ref{wij})).
%Интеграл в~(\ref{onq}) несложно вычислить, если представить его в виде
The integral in Eq.~(\ref{onq}) is easy to calculate if one represents it as
\begin{eqnarray}
\label{int4}
-\frac{\partial}{\partial q_i} \frac{\partial}{\partial q_j} \int d^3 r \frac{e^{i{\bf qr}}}{4\pi r^2}\frac{w^2(r)}{r^2}\hat S_{12}(\hat{\bf r}; {\bf S}^{(2)},{\bf S}^{(2)}) = \nonumber \\
\frac{\partial}{\partial q_i} \frac{\partial}{\partial q_j} \int dr w^2(r)\frac{j_2(qr)}{(qr)^2}\,\hat S_{12}({\bf q}; {\bf S}^{(2)},{\bf S}^{(2)}),
\end{eqnarray}
where
$\hat S_{12}({\bf q}; {\bf S}^{(2)},{\bf S}^{(2)}) = 3({\bf S}^{(2)}\cdot {\bf q})^2 - 2q^2$.
After calculating the integral, we get four terms proportional to symmetric tensors
$\delta_{ij}$, $\hat q_i \hat q_j$, ${\bf S}^{(2)}_i \hat q_j + {\bf S}^{(2)}_j \hat q_i$ and $\{S^{(2)}_i,S^{(2)}_j\}$.
The first two terms are vanishing when multiplied by the vector ${\bf W}'^n_{ij}$, and the third turns to zero when multiplied by its $\hat q$-component, which is involved in calculating the TVPC signal
(when taking non-diagonal spin matrix elements from the product
${\bf S}^{(1)}\cdot{\bf W}'^n_{ij}$).
%%YU
%--
%(see (\ref{g12m}) -
%странная ссылка на эту формулу).
%%MP убрала ссылку
Thus, the contribution to the TVPC signal is given only by a term proportional to
$\{S^{(2)}_i,S^{(2)}_j\}$,
which is obtained by differentiating the operator
$\hat S_{12}({\bf q}; {\bf S}^{(2)},{\bf S}^{(2)})$ in (\ref{int4}).
Rewriting $\frac{j_2(qr)}{(qr)^2}$ via a linear combination of spherical Bessel functions
$j_n(qr), n=0,2,4$,
we get the following contribution to the TVPC signal from the last term in Eq.~(\ref{onq}):
\begin{equation}
{\bf S}^{(1)}\cdot{\bf W}'^n_{ij}\{S^{(2)}_i,S^{(2)}_j\}\left[\frac{3}{5}S_0^{(2)}(q) - \frac{6\sqrt{2}}{7}S_2^{(2)}(q) + \frac{18}{35}S_4^{(2)}(q)\right],
\end{equation}
where the form factor $S_4^{(2)}(q)$ is defined  in (\ref{s02}).

When integrating the operator $\hat\Omega^{2 n}({\bf q})$ with the factor
$e^{i{\bf q}\bm{\rho}}$ by the coordinates of the nucleons in the incident deuteron, we obtain an expression similar to (\ref{onq}),
with the replacement ${\bf W}'^n_{ij} \to {\bf W}^n_{ij}$ and ${\bf S}^{(1)} \leftrightarrow {\bf S}^{(2)}$.
For abnormal scattering, we also obtain similar expressions
(with the replacement $Z_0 \to Z(q)$)
when integrating $\hat\Omega^{2 a}_q({\bf q})$ with the factor
$e^{-i{\bf q}\bm{\rho}}$
over the nucleon coordinates in the beam,
and $\hat\Omega'^{2a}_q({\bf q})$ with the factor $e^{i{\bf qr}}$
over the nucleon coordinates in the target.

Now, to get the amplitude $M_{\rm TVPC}(0)$ in the double-scattering approximation, we take the sum of all expressions of the form (\ref{onq}) for normal and abnormal scattering, integrate by the momentum ${\bf q}$ and multiply by a factor $\frac{i}{2\pi^{3/2}}$ (see Eq. (\ref{mg2})).
From the resulting operator, we calculate the spin matrix elements~(\ref{g12m}) necessary to find the TVPC amplitudes
$g_i$, $i=1,2$.
To do this, we use the following relations:
\begin{eqnarray}
\label{m26s}
<-1,1|{\bf S}^{(1)}\cdot\hat{\bf q}\{{\bf S}^{(2)}\cdot\hat{\bf n},{\bf S}^{(2)}\cdot\hat{\bf k}\}|0,0> \!\!&=&\!\! \nonumber \\
-\!<\!-1,1|{\bf S}^{(1)}\!\cdot\!\hat{\bf q}\{[{\bf S}^{(2)}\!\times\! \hat{\bf q}]\!\cdot\! \hat{\bf n},[{\bf S}^{(2)}\!\times\! \hat{\bf q}]\!\cdot\! \hat{\bf k}\}|0,0\!> \!\!&=&\!\! -\frac{i}{2}, \nonumber \\
<1,0|{\bf S}^{(1)}\cdot\hat{\bf q}\{{\bf S}^{(2)}\cdot\hat{\bf n},{\bf S}^{(2)}\cdot\hat{\bf k}\}|0,1> \!\!&=&\!\! \nonumber \\
-\!<\!1,0|{\bf S}^{(1)}\!\cdot\!\hat{\bf q}\{[{\bf S}^{(2)}\!\times\! \hat{\bf q}]\!\cdot\! \hat{\bf n},[{\bf S}^{(2)}\!\times\! \hat{\bf q}]\!\cdot\! \hat{\bf k}\}|0,1\!> \!\!&=&\!\! \frac{i}{2}.
\end{eqnarray}
When replacing ${\bf S}^{(1)} \leftrightarrow {\bf S}^{(2)}$, both matrix elements in~(\ref{m26s}) turn out to be the same and equal to $-\frac{i}{2}$.

As a result, we get the formulas (\ref{g12}).

%\bibliography{pu-ddmm}%
%\bibliography{pu-ddm5c}%
\bibliography{pu-ddm7c}%
%\bibliography{Introd}% Produces the bibliography via BibTeX.

\end{document}